\pdfoutput=1
\documentclass[fleqn,superscriptaddress]{revtex4}
\usepackage[utf8x]{inputenc}
 \usepackage{epsfig}
\usepackage{amssymb}
 \usepackage{amsthm}
 \usepackage{color}
\usepackage{amsmath}
\usepackage{empheq}

\newcommand{\p}{\partial}
\newcommand{\bea}{\begin{eqnarray}}
\newcommand{\eea}{\end{eqnarray}}
\newcommand{\be}{\begin{equation}}
\newcommand{\ee}{\end{equation}}
\newcommand{\bse}{\begin{subequations}}
\newcommand{\ese}{\end{subequations}}

\renewcommand{\l}{\ell}

\newcommand{\comment}[1]{}
\renewcommand{\ss}[1]{_{\hbox{\tiny #1}}}

\newcommand{\mb}[1]{\mathbf{#1}}

\begin{document}





\title{Coarsening dynamics at unstable crystal surfaces}


\author{Paolo Politi}

\address{Istituto dei Sistemi Complessi, Consiglio Nazionale delle Ricerche,
via Madonna del Piano 10, 50019 Sesto Fiorentino (Italy)\\
paolo.politi@cnr.it}

\begin{abstract}
In this paper we focus on crystal surfaces led out of equilibrium by a growth or erosion process.
As a consequence of that the surface may undergo morphological instabilities and develop a distinct
structure: ondulations, mounds or pyramids, bunches of steps, ripples.
The typical size of the emergent pattern may be fixed or it may increase in time through a coarsening process
which in turn may last forever or it may be interrupted at some relevant length scale.
We study dynamics in three different cases, 
stressing the main physical ingredients and the main features of coarsening:
a kinetic instability, an energetic instability, and an athermal instability. 
\end{abstract}

\maketitle






\section{Preamble}

This paper appears in a special issue of Comptes Rendus Physique devoted to
coarsening dynamics. My contribution focuses on coarsening processes occurring
at the surface of a crystal as a result of an instability.
The meaning of coarsening in this context is clearly exemplified
by the images (a1-a6) in Fig.~1. 
Here we provide Scanning Tunneling Microscope (STM) images of a growing
Cu/Cu(100) surface at different times (the number of deposited monolayers (ML)
is proportional to time and it is indicated on the bottom right of each image).
As you can see, the crystal surface does not growth flat and it develops a clear
mound structure whose typical size increases in time.
Images (c1-c4) in the same figure refer to a similar growth process of Pt/Pt(111).
In this case coarsening stops after a while.

There are so many unstable growth processes at a crystal surface that it is not
possible to cover all them in this short review.
For this reason I will treat a personal shortlist of phenomena. Also, I will not
try to report all relevant bibliography, because I am more interested 
to develop general considerations and to highlight some aspects.

I have chosen three main topics with different features:
i)~The kinetic instability of a growing high symmetry surface.
It takes place because the growth process is too fast and the
surface cannot relax to its equilibrium shape (a flat surface
in this case).
ii)~The energetic instability of a surface in heteroepitaxy.
Here, dynamics is driven by the minimization of the system free energy.
iii)~The nonthermal instability of a surface under ion beam erosion.
In this case surface morphology changes under the action of an incident beam
of energetic ions.

The first topic on the kinetic instability, which is introduced in Sec.~\ref{sec_ph} and discussed
in Sec.~\ref{sec_hs}, is the most extended one because
it has long been studied in {\it specific} relation to coarsening.
In fact, experimental studies of this phenomenon show all possible scenarios.
Here above we have already mentioned perpetual coarsening and interrupted
coarsening. The former, illustrated in Fig.~1(a1-a6), means that the size
of the emergent pattern increases in time forever. The latter, shown in
Fig.~1(c1-c4), means that coarsening stops when the size reaches some
special length scale, not depending on boundary conditions.
Finally, it is also possible that the size of the pattern keeps constant 
in time (no coarsening). This case is shown in images (b1-b2) of Fig.~1,
where atomic steps of a growing Cu surface meander with a typical lengthscale
which does not growth in time.

It may seem amazing we mention ``no coarsening'' as a coarsening scenario,
but all these scenarios can be understood within the same picture and model.
This remark also allows to introduce a second, important reason
for which topic (i) is discussed more widely:
we have a much deeper theoretical comprehension of its coarsening behaviour
than for other topics. This is possible because the 
scientific community has agreed on the special relevance of given
growth equations and many studies have been devoted to them. 

In Sec.~\ref{sec_el} we discuss an energetic instability occurring in heteroepitaxy.
In this case we can imagine to deposit a semiconductor (e.g., Ge) on top of
a different one (e.g., Si). Since the two elements have different lattice
constants, the adsorbate must squeeze or strectch in order to growth coherently with the
substrate. This fact leads to an accumulation of elastic energy which
then must be released. A possible way out is a
phenomenon with relevant applications:
the formation of quantum dots through a self-organized growth process.
This is illustrated in images (d1-d4) of Fig.~1, where a suitable combination
of Si and Ge is deposited on top of Si(100).
In this case the source of the instability is the minimization of the 
elastic energy, whose nonlocal character makes much more involved
a theoretical description. It is not surprising that there is no simple equation
able to describe the main features of the dynamics.

Finally, in Sec.~\ref{sec_er} we study an athermal instability,
due to the erosion of a crystal surface with an incident flux of
high energy ions. The resulting morphology is shown in images
(e1-e4) of Fig.~1 for the case of Si(001) under the exposure to Kr$^+$.
In ion beam erosion the elementary process changing surface morphology
is the ejection of surface atoms because of the impact of
a high energy ion. This process is not thermally activated,
as opposed to surface processes driving the dynamics of the systems
in topics (i) and (ii).
Theoretical descriptions of the evolution of a surface under ion beam erosion are very
phenomenological and the experimental picture does not seem
to be entirely clear.

I hope both people working and not working in this field find something useful. 
The non specialist should understand the basic physical ingredients of the
phenomena we are going to discuss and get a general idea of them. 
The specialist might find of some interest the focus on coarsening in different contexts.
Some aspects of an unstable crystal surface closely resemble phase separation 
while other aspects fall into the category of pattern formation.
For this reason the reader may find some connections with the contributions by
Leticia Cugliandolo and by Alexander A. Nepomnyashchy in this issue. 

\section{The physical picture}
\label{sec_ph}

A crystal surface at thermal equilibrium  undergoes the well known roughening
transition, according to which above the temperature $T=T_R$ the system is characterized by diverging  height
fluctuations. Below $T_R$ a crystal surface is characterized by orientations of minimal
free energy. Therefore, quenching the system to $T<T_R$, an unstable surface splits into regions
of different orientations (facets) whose size increases in time through a process of phase separation
where the local slope is the order parameter~\cite{Stewart_Goldenfeld,Liu_Metiu}.
In the following we are not dealing with such process, 
but we are mostly speaking about thermal systems which are kept out of equilibrium
by a flux of atoms. 

Surface dynamics has two special features with respect to bulk dynamics:
it can be more easily imaged and it is faster.
The former feature is trivial. The latter is due to the fewer number
of neighbouring atoms, which makes thermally activated processes more likely.
The ensemble of atomistic processes occurring at a crystal surface is
schematically depicted in Fig.~2, where two main objects stand out, steps and adatoms.
Adatoms (adsorbed atoms) are provided by the incoming flux and by thermal detachment from steps.
They diffuse until they (re)attach to steps or go (back) to the vapour phase.
If the surface is cut along a high symmetry orientation, e.g. the orientation (100) of a
cubic crystal, the perfect surface does not contain steps, which are created
by the nucleation and aggregation processes, when adatoms stick together.
Conversely, if the surface is not cut along a high symmetry orientation, then
steps are naturally present.

The key point is that step dynamics is usually much slower than adatom dynamics.
This may be a drawback for atomistic simulations but it is useful for analytical treatments.
In fact, simulating a system where very different time scales are relevant is problematic.
On the other hand, different time scales may allow to adopt quasi static approximations,
where the dynamics of certain (fast) variables is {\it slaved} to the instantaneous value of
other (slow) variables.
In practice, adatom diffusion can be studied assuming steps are fixed, then step motion
is determined via the average diffusion current attaching to or detaching from steps.

\begin{figure}
\begin{minipage}{0.74\textwidth}
\includegraphics[width=\textwidth]{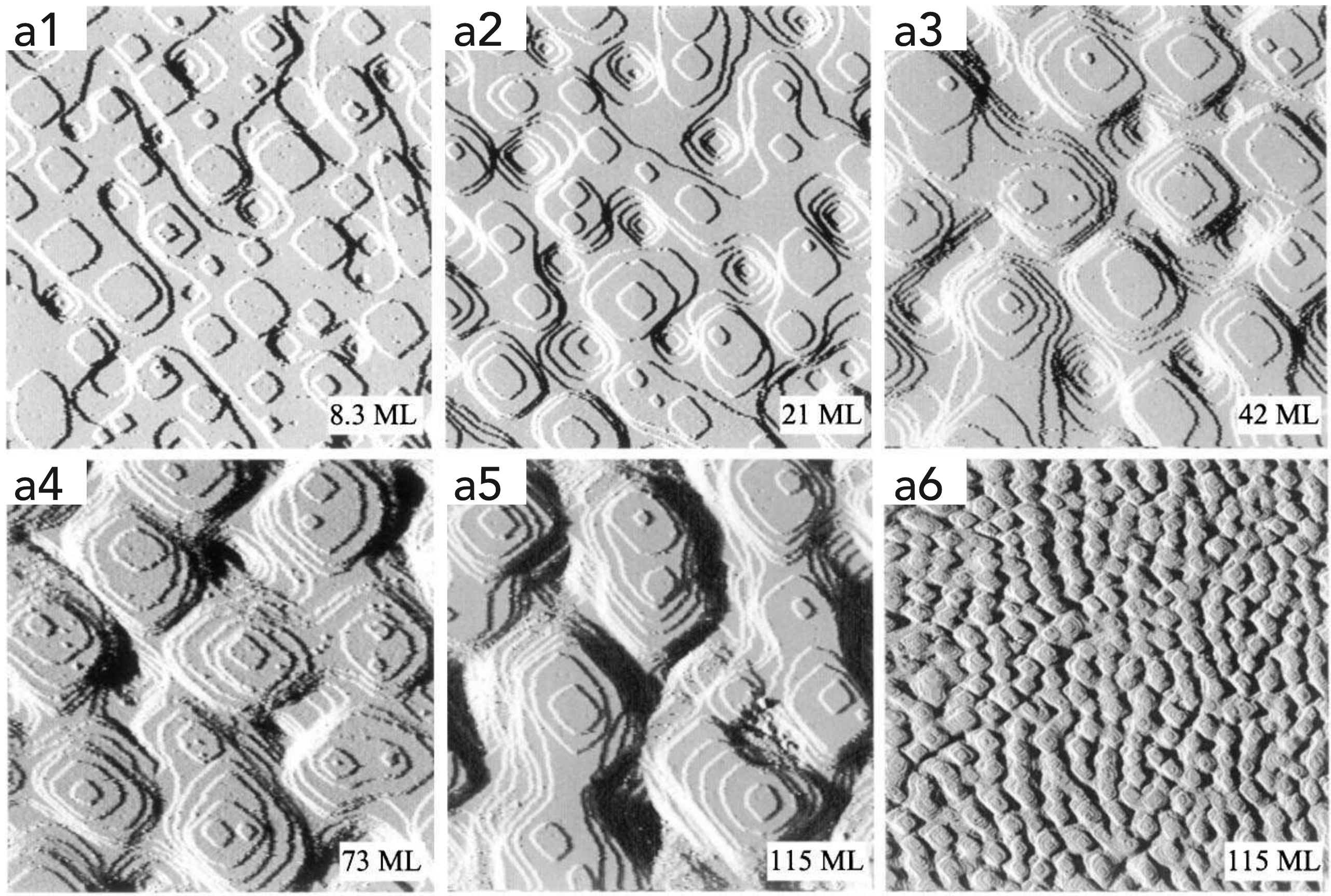}
\end{minipage}
\begin{minipage}{0.24\textwidth}
\includegraphics[width=\textwidth]{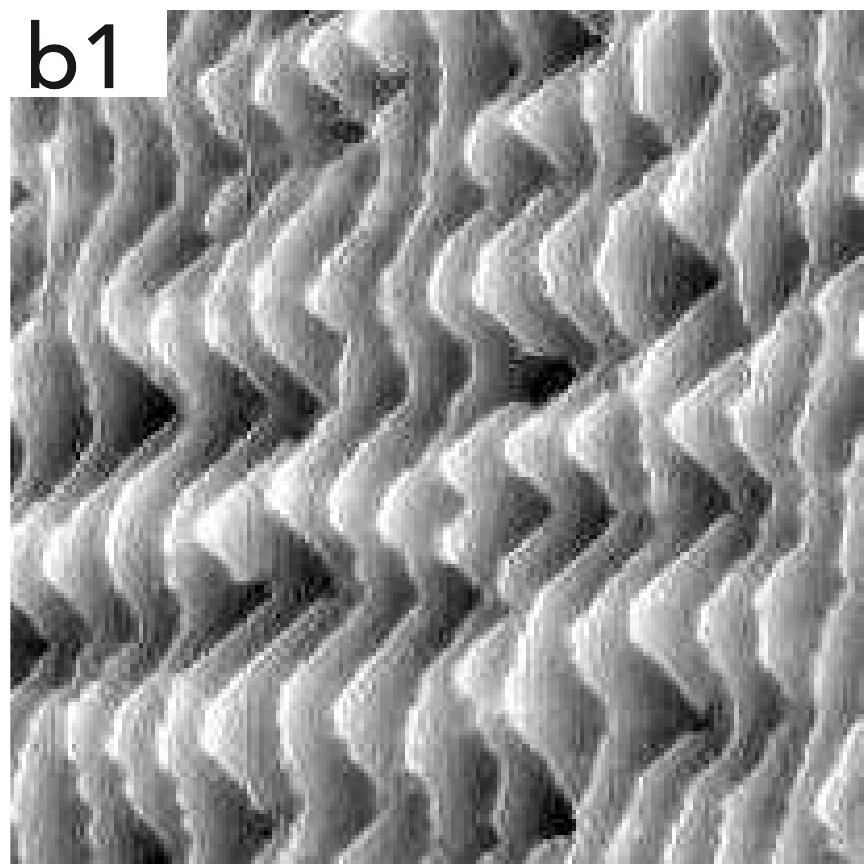}
\includegraphics[width=\textwidth]{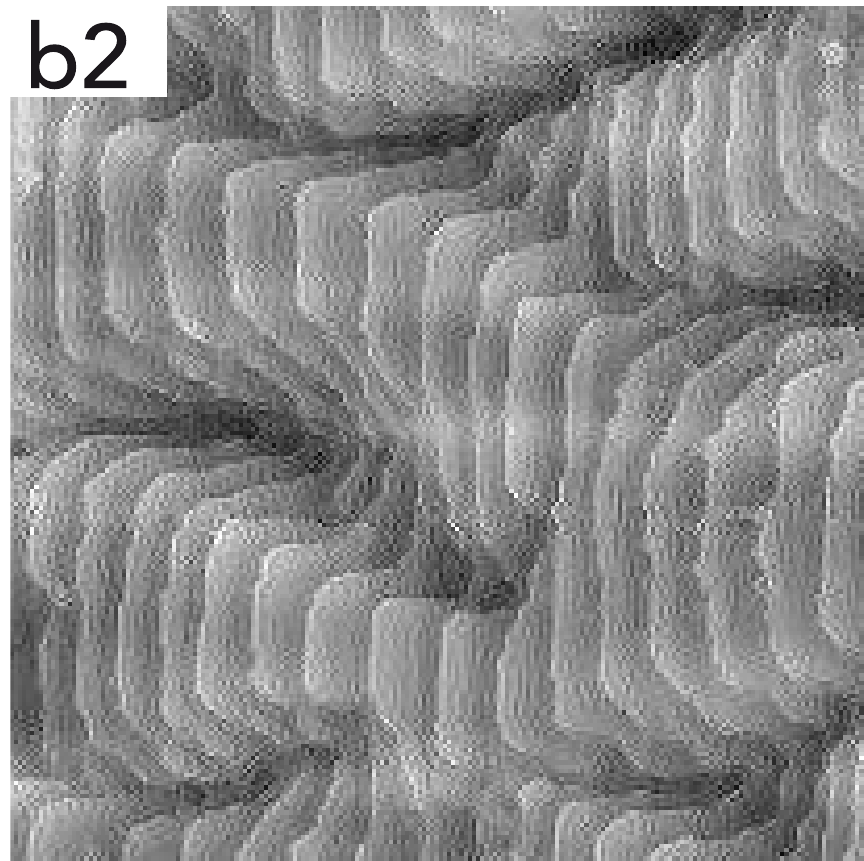}
\end{minipage}

\includegraphics[width=0.24\textwidth]{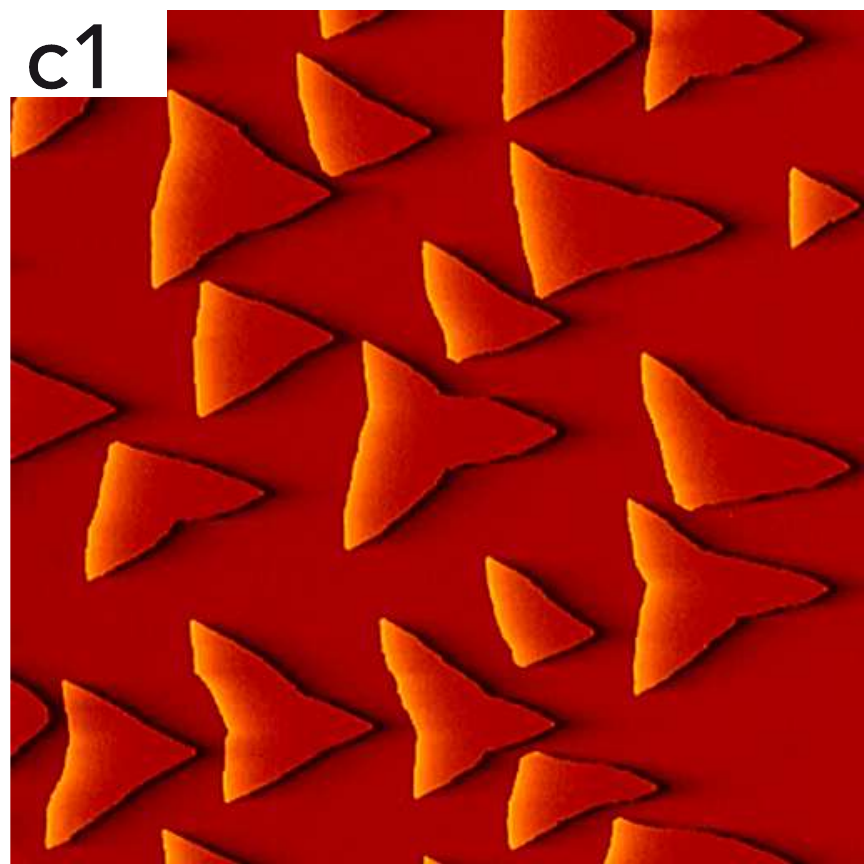}
\includegraphics[width=0.24\textwidth]{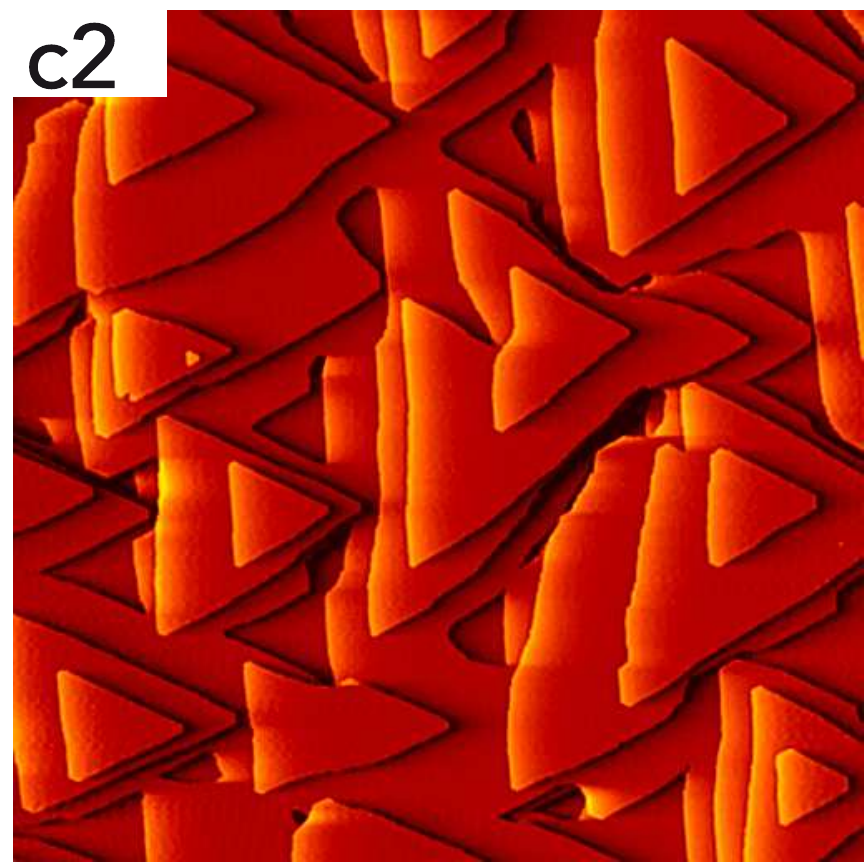}
\includegraphics[width=0.24\textwidth]{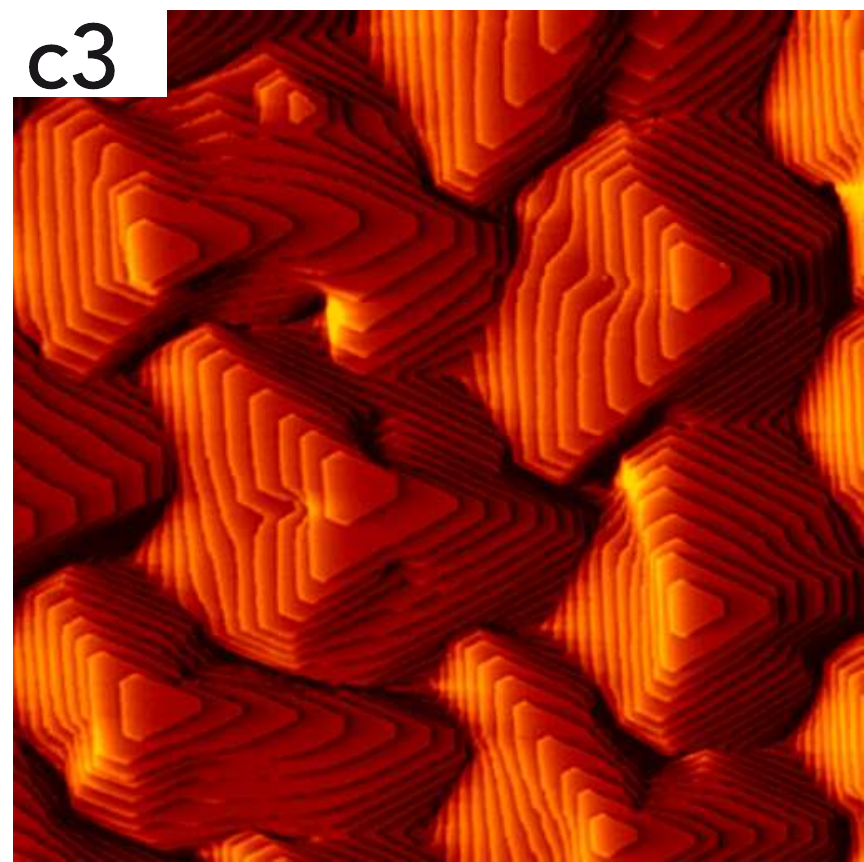}
\includegraphics[width=0.24\textwidth]{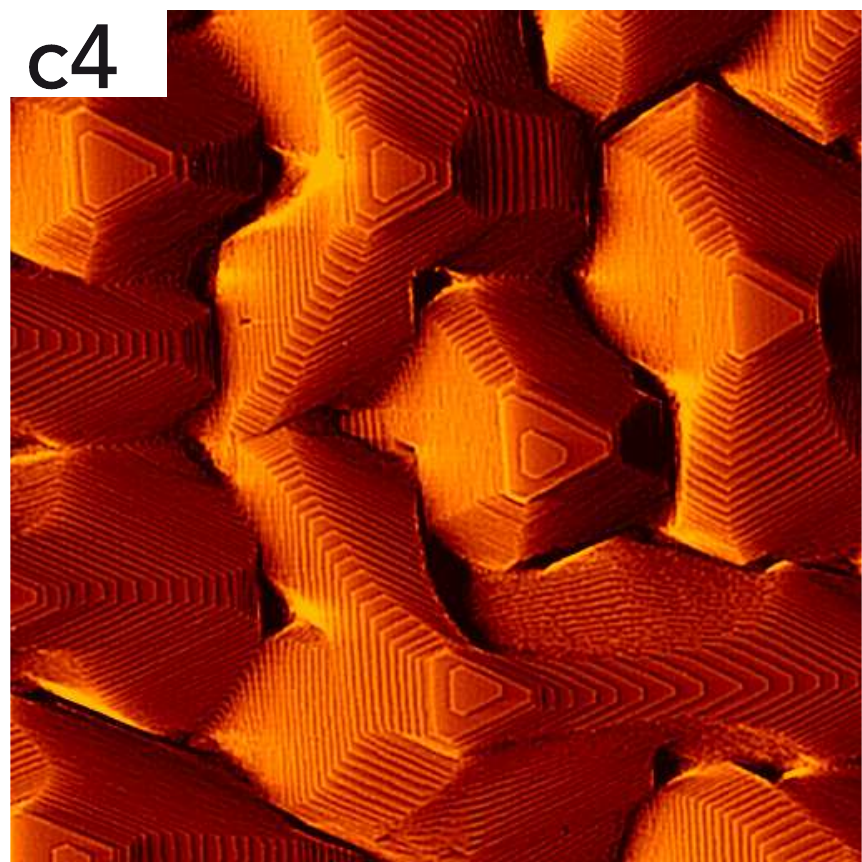}

\includegraphics[width=0.24\textwidth]{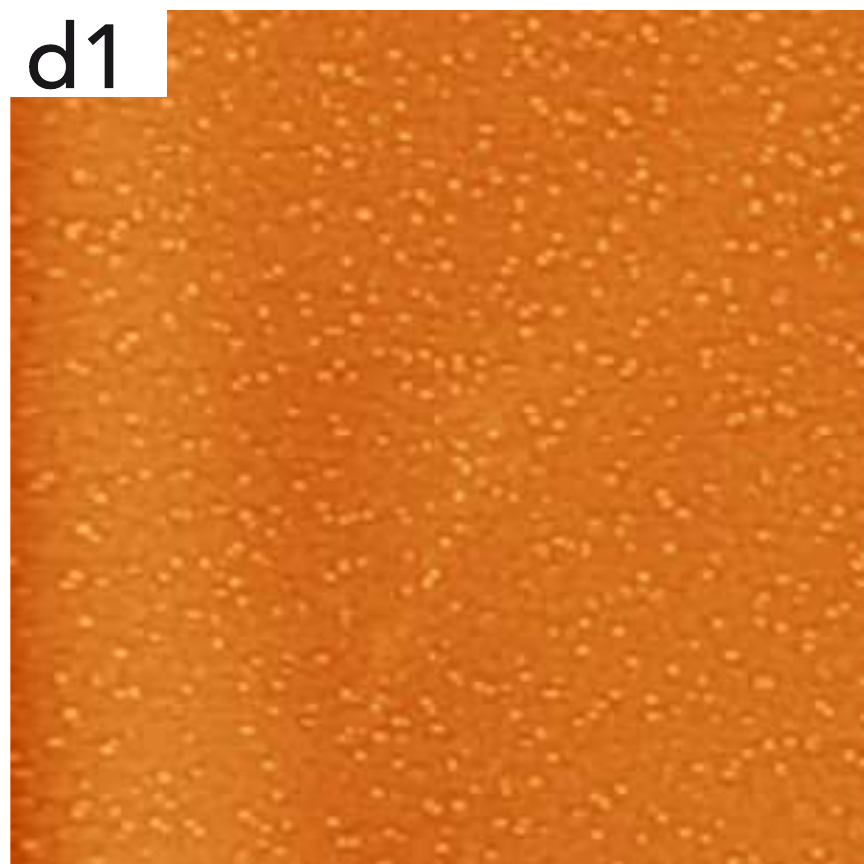}
\includegraphics[width=0.24\textwidth]{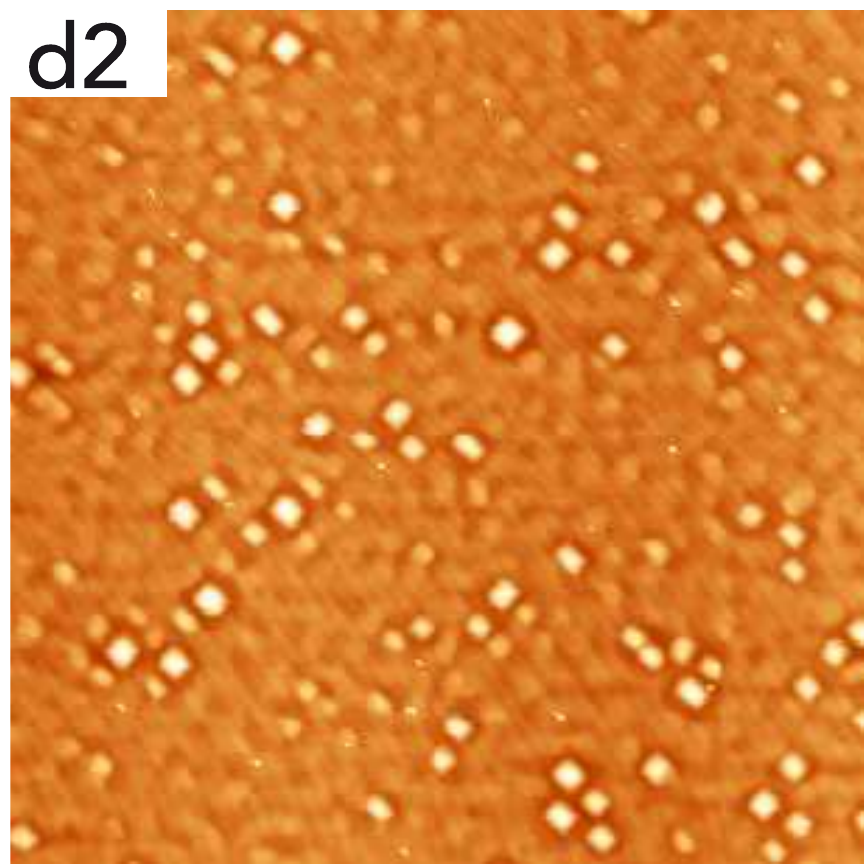}
\includegraphics[width=0.24\textwidth]{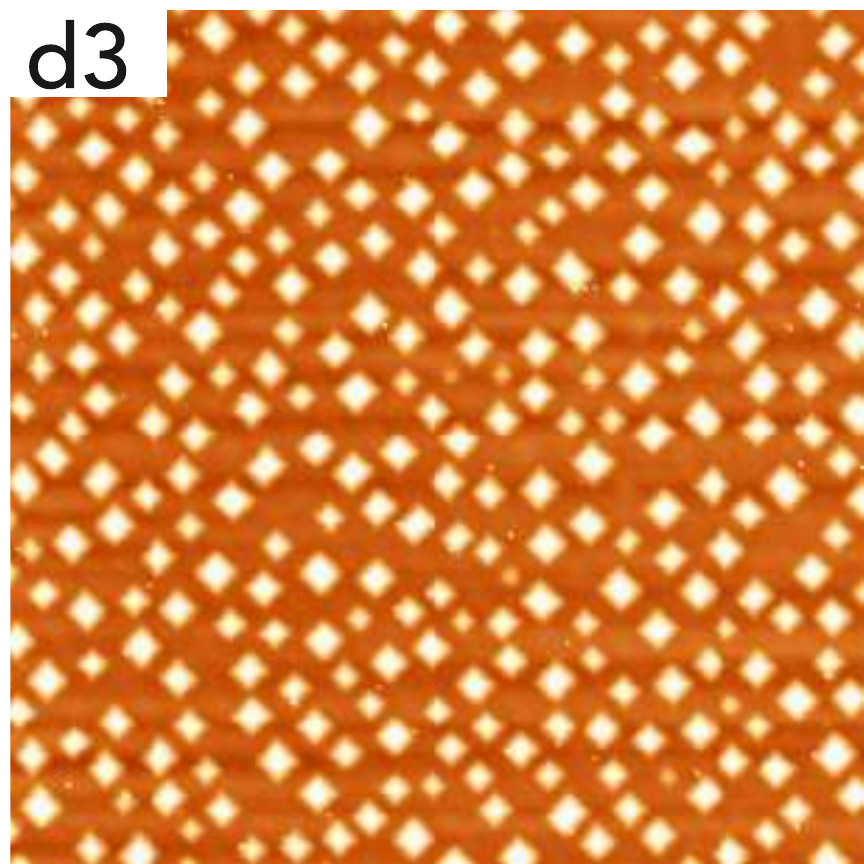}
\includegraphics[width=0.24\textwidth]{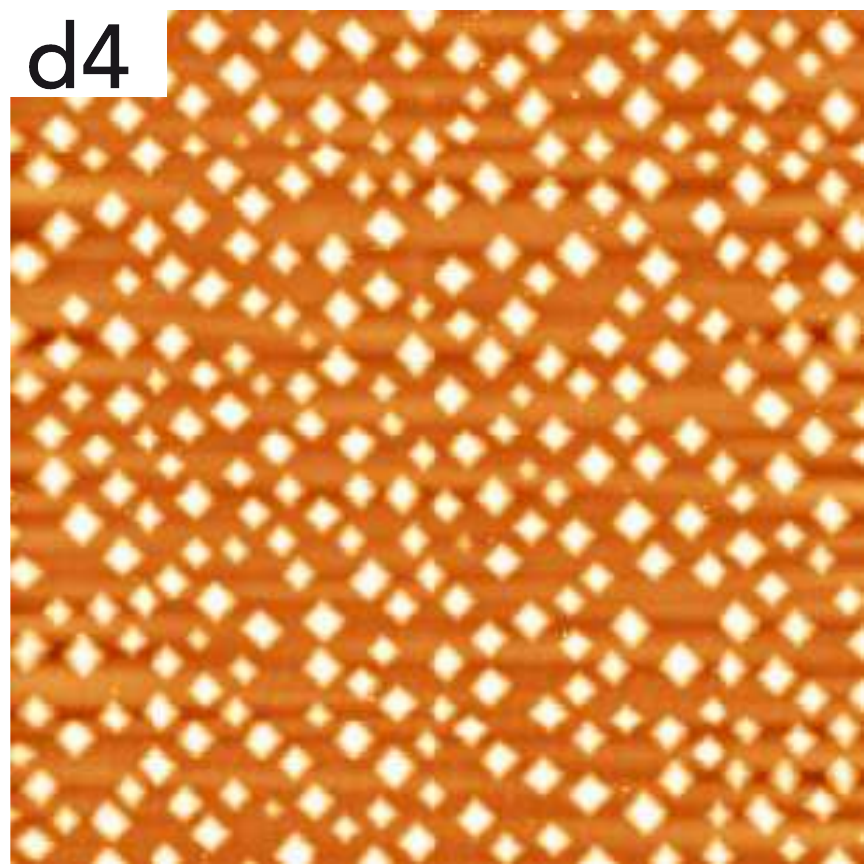}

\includegraphics[width=0.24\textwidth]{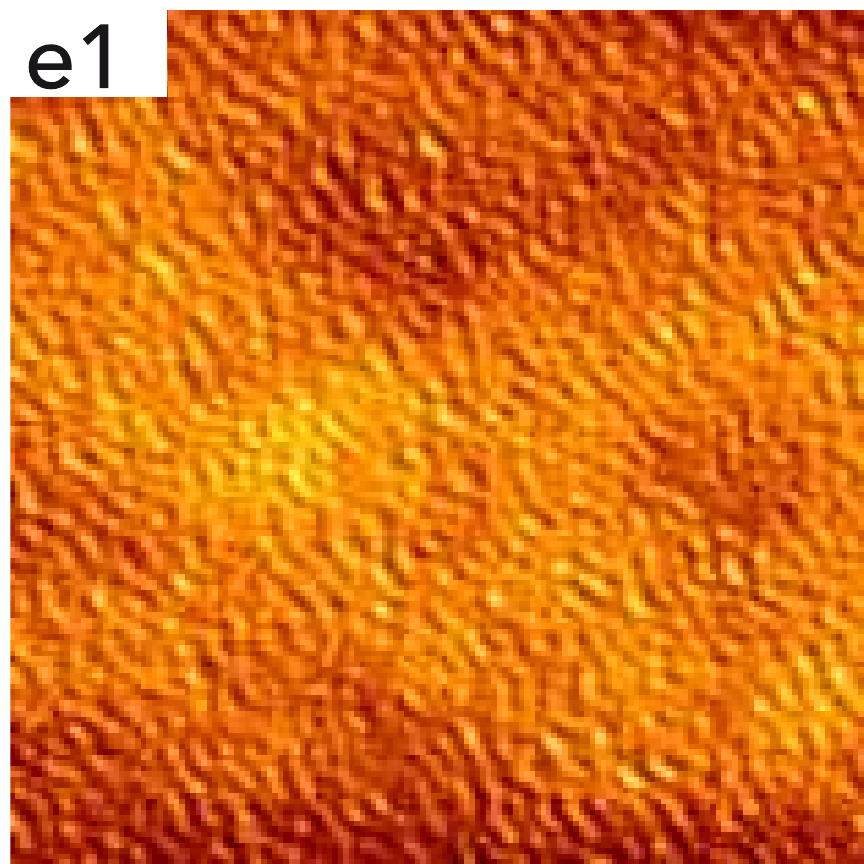}
\includegraphics[width=0.24\textwidth]{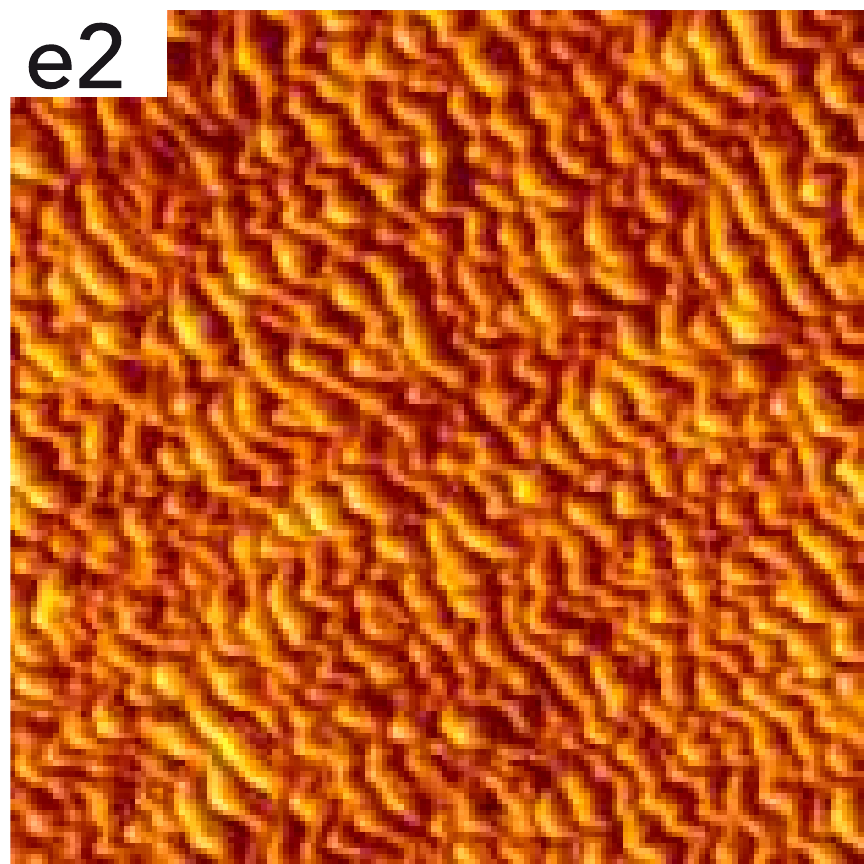}
\includegraphics[width=0.24\textwidth]{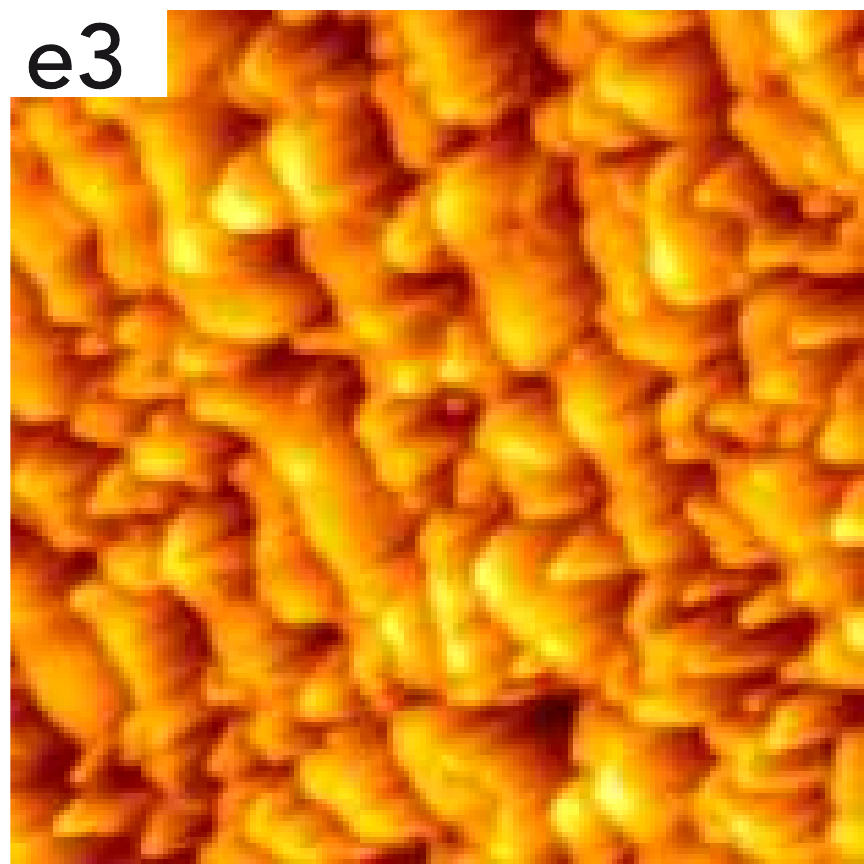}
\includegraphics[width=0.24\textwidth]{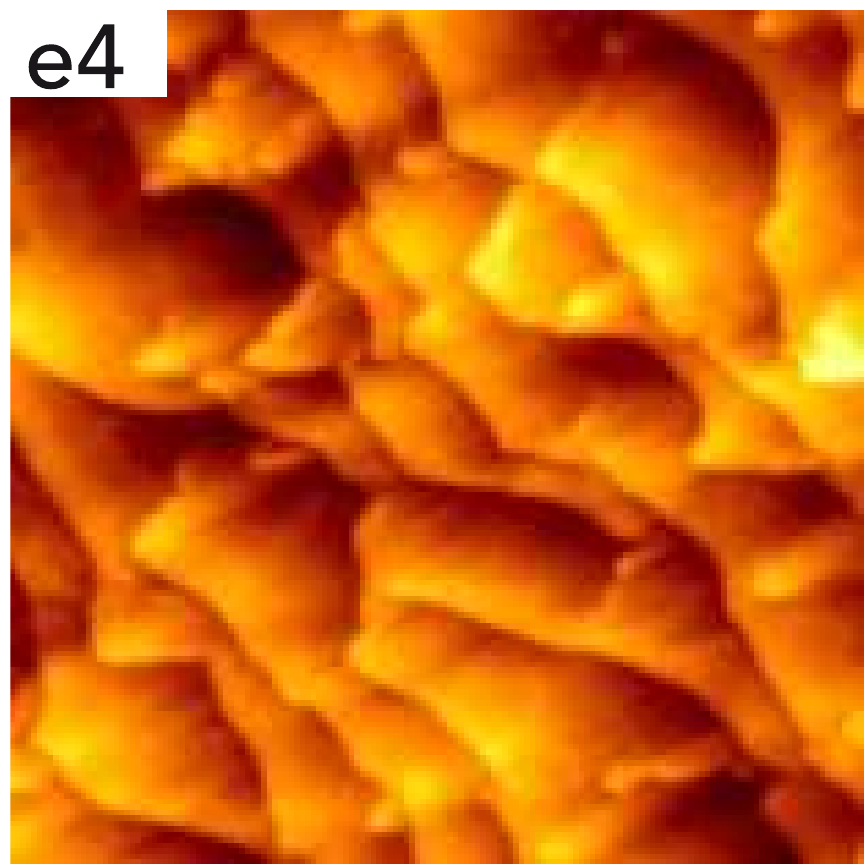}
\caption{
(a1-a6) STM images of Cu/Cu(100) at various thicknesses. The area is $1000\times 1000$\AA (a6 is
$5000\times 5000$\AA). Taken from Ref.~\cite{Zuo_Wendelken}.
(b1-b2) STM images of (b1) Cu(0 2 24) and (b2) Cu(1 1 17). The area is $40\times 40$nm.
Taken from Ref.~\cite{Douillard}.
(c1-c4) STM images of Pt/Pt(111) at (c1) 0.35 ML, (c2) 3 ML, (c3) 12 ML, (c4) 90 ML.
The area is $2900\times 2900$\AA. Taken from Ref.~\cite{Kalff}.
(d1-d4) AFM images of Si$_{0.75}$Ge$_{0.25}$/Si(100) (d1) as grown, (d2) after 1.5h of annealing, 
(d3) after 18h of anenaling, and (d4) after 54h of annealing. 
The area is $5\times 5\mu$m$^2$, the vertical scale is 25nm. Courtesy of Isabelle Berbezier.
(e1-e4) STM images of Si(001) after 2 keV Kr$^+$ ion exposure and ion fluences of 
(e1) $10^{20}$, (e2) $3\times 10^{20}$, (e3) $10^{21}$, (e4) $3\times 10^{21}$ ions/m$^2$.
Taken from Ref.~\cite{Engler}.
}
\label{fig_exp}
\end{figure}

\begin{figure}
\begin{center}
\includegraphics[width=0.8\textwidth]{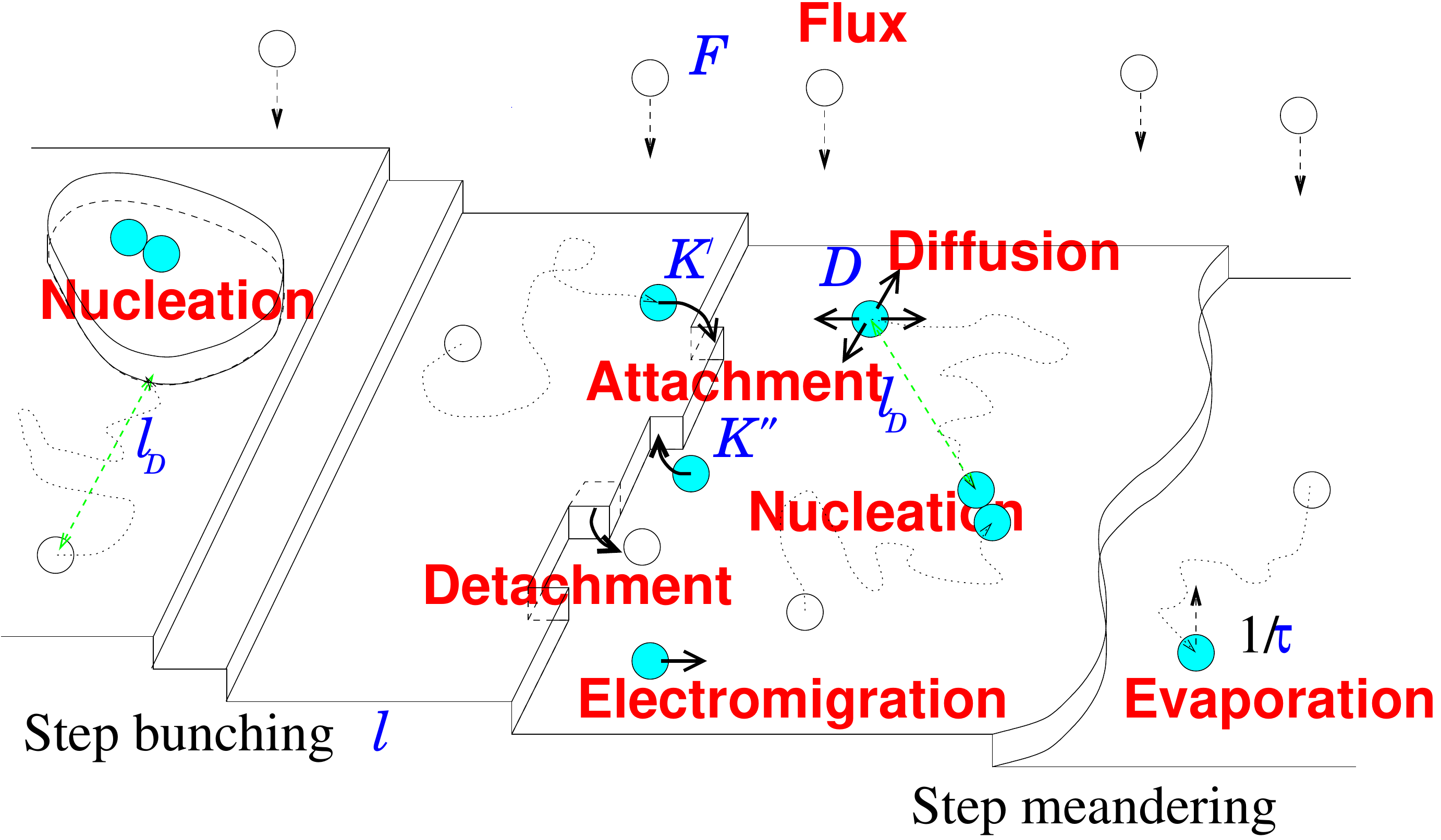}
\end{center}
\caption{Schematics of basic elementary processes occurring at a crystal surface.
Adatoms (empty circles) are provided by the incoming flux (at rate $F$) and by detachment from steps. 
Once an adatom is on a terrace, it may diffuse (at rate $D$), it may evaporate (at rate $\tau^{-1}$), 
it may attach to a step (at rates $K',K''$) or it may
encounter another adatom and nucleate a new terrace.
Steps may be preexisting (long straight or meandering steps) or may be created by nucleation
(step encircling the nucleated island).
Two lengths are specially important during growth: the distance $\ell$ between preexisting steps and the
diffusion length $\ell_D$, i.e. the distance traveled by an adatom before nucleating a new terrace or being
incorporated in a terrace. If $\ell\gg\ell\ss{D}$, the surface behaves as a high symmetry surface and
preexisting steps play no role. In the opposite limit, $\ell\ll\ell\ss{D}$, dynamics is dominated by step-flow growth
and nucleation of new terrace is absent.
Three main instabilities may occur: step-bunching (a train of equally spaced steps is unstable
against fluctuations of their distance); step-meanderings (a straight step is unstable);
mound formation (islands are nucleated at a rate larger than their coalescence, so that the flat surface
is unstable).
}
\label{fig_mbe}
\end{figure}

\begin{figure}
\begin{center}
\includegraphics[width=0.38\textwidth]{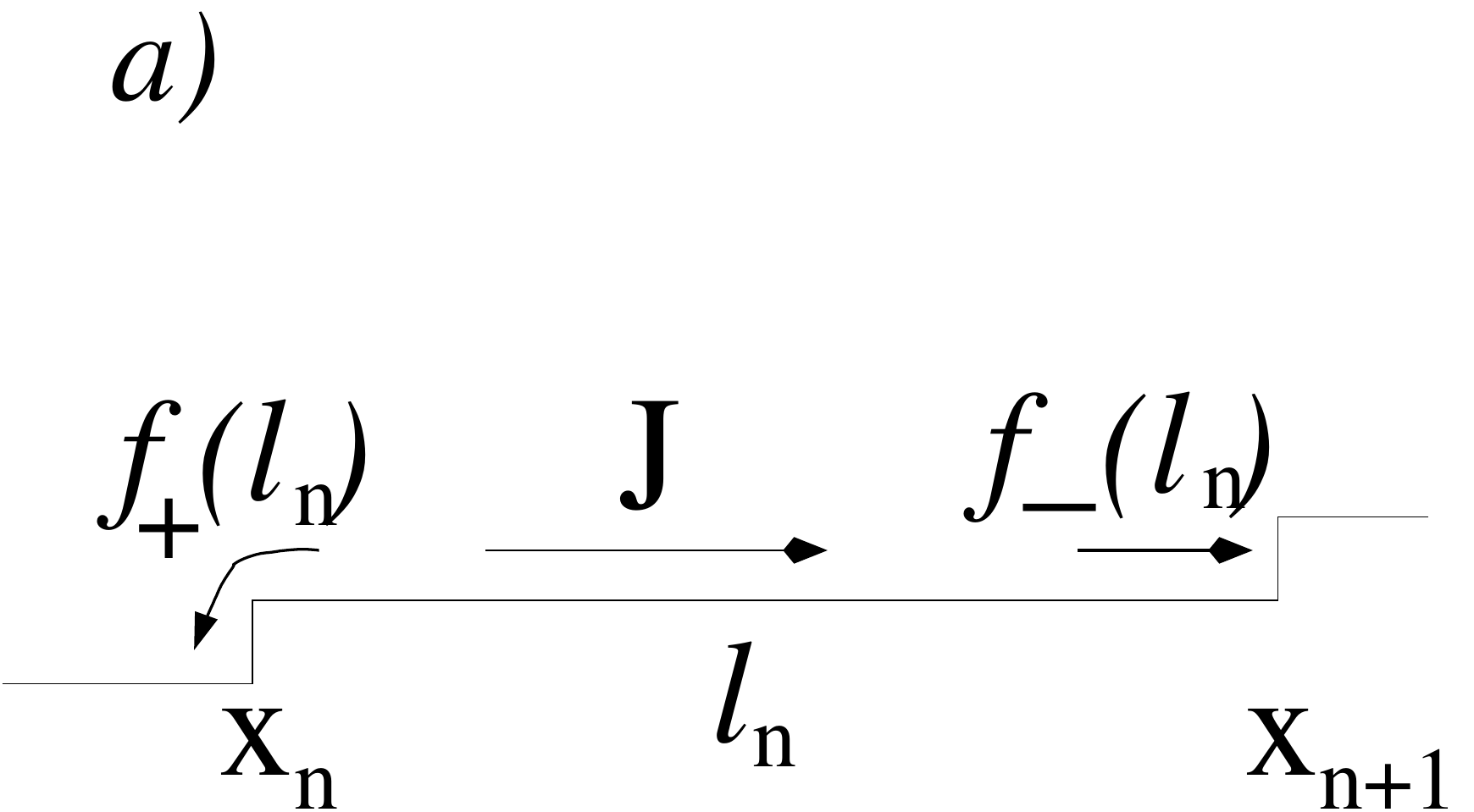}
\hfill
\includegraphics[width=0.5\textwidth]{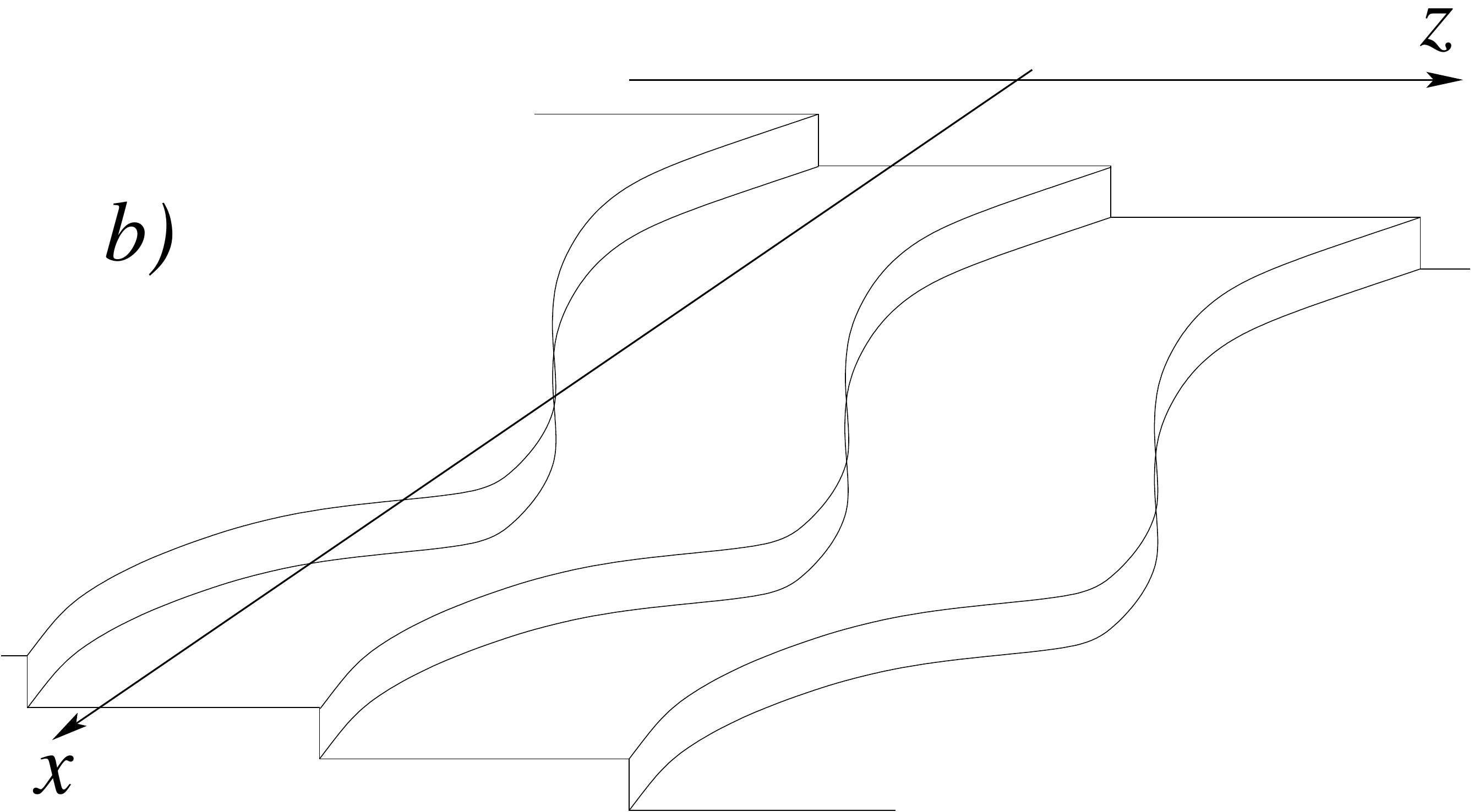}
\end{center}
\caption{a)~Adatoms may attach to or detach from the upper and lower step.
The quantity $f_+(\ell)$ ($f_-(\ell)$) represent the net balance of 
the exchange of adatoms between a step and the upper (lower) terrace
of size $\l$. A positive $f_\pm$ means that more adatoms attach than detach.
The current $J$ is defined as $J=f_-(\ell)-f_+(\ell)$.
b)~A train of equally spaced steps which meander in phase. This type of instability
allows to reduce 2D step dynamics to an effective 1D model for the variable $z(x,t)$.}
\label{fig_vicinal}
\end{figure}

The true origin of thermal instabilities we are going to discuss is the existence of 
an asymmetric diffusion current between steps, which may be preexisting (not high simmetry surface) 
or be created by the dynamics itself (high symmetry surface). 
Such net current can have an intrinsic or an extrinsic origin. Surface reconstruction,
asymmetry in step attachment/detachment rates ($K'\ne K''$ in Fig.~2), 
and step-step interaction are intrinsic motivations.
Impurities and an applied electric current are among extrinsic motivations.
Let us explain with a simple one-dimensional model why a current asymmetry may produce an instability,
see Fig.~\ref{fig_vicinal}a.

The speed of a step depends on the net balance between attaching and detaching adatoms.
It is useful to separate the balance of exchanging atoms with the upper terrace, $f_+$,
from the balance of exchanging atoms with the lower terrace, $f_-$, because a net
current $J$ appears when they are different: $J(\l)=f_-(\l)-f_+(\l)$. 
Each contribution depends on the size $\ell$ of the terrace, of course.
Making reference to Fig.~\ref{fig_vicinal}a, we can write
\be
\dot x_n = f_+(\ell_n) + f_-(\ell_{n-1}) \qquad
\mbox{and} \qquad
\dot\ell_n = \dot x_{n+1}-\dot x_n .
\label{eq_vn}
\ee
It is useful to separate contributions in a symmetric ($s(\l)$) and an antisymmetric ($a(\l)$)  part,
\be
f_\pm(\ell) = s(\ell)\pm a(\ell),
\ee
so that $J(\l)=f_-(\l)-f_+(\l)=-2a(\l)$. Replacing in Eq.~(\ref{eq_vn}) we get
\be
\dot \ell_n =[s(\ell_{n+1})-s(\ell_{n-1})]
+ [a(\ell_{n+1})+a(\ell_{n-1})-2a(\ell_n)] ,
\label{eq_ell}
\ee
which immediately clarifies the different role of the two parts, the symmetric
one entering with a first derivative, the antisymmetric one entering via a second derivative.
We can now perform a linear stability analysis of the uniform train of steps, writing
\be
\ell_n(t) = \ell +\epsilon_n(t) =\ell +\epsilon_0 \exp(iqn+\sigma t)
\label{eq_lnl}
\ee
where a positive real part of $\sigma(q)$ signals an instability. 
Replacing Eq.~(\ref{eq_lnl}) in (\ref{eq_ell}) we finally obtain
\be
\sigma(q) = (1-\cos q)J'(\ell) + i\,\big( 2 s'(\ell)\sin q \big) ,
\ee
so that
\be
\mbox{instability}\quad \longleftrightarrow \quad J'(\ell) > 0 .
\ee
Strictly speaking, this analysis is valid for a surface looking like a staircase,
where all steps are of the same type: ascending or descending steps. 
This is the case of images (b1-b2) in Fig.~1, which represent a {\it vicinal}
surface of copper.\footnote{It is called vicinal because its Miller indexes are close
(vicinal) to those of a high symmetry surface.}
The nucleation and aggregation processes determine ascending and descending steps, producing
a morphology like the one you can see in Figs.~1(a1-a6)
and Figs.~1(c1-c4). In this case steps are continuosly created and destroyed and 
a mesoscopic continuum description, in terms of the local height $z(\mb{x},t)$, is preferable.
This is the approach used in the next Section.

\section{High symmetry orientations}
\label{sec_hs}

In very general terms, the local velocity of the surface, $\p_t z(\mb{x},t)$, is a function
of the surface profile. If we neglect long-range forces (essentially elasticty, in this context)
such depedence is limited to $z(\mb{x},t)$ and to spatial derivatives of various order.
In the case the deposited material is the same of the substrate (homoepitaxial growth),
translational invariance in the growth direction does not allow for an explicit dependence on $z$.
Finally, if desorption is negligible dynamics ``reduces" to an ensemble of surface processes
which conserve matter. Therefore, if $z(\mb{x},t)$ is the local height with respect to the average 
height of the surface, its time derivative must satisfy the continuity equation,
\be
\p_t z(\mb{x},t) =-\nabla\cdot\mb{j}(\nabla z,\nabla^2z,\dots) . \label{eq_2d}
\ee
The rest of this Section will be devoted to introduce a couple of relevant equations, one 
in $D=1$ and one in $D=2$, and to discuss what we know about them. 
The case 1D, simpler than the case 2D, will also serve to introduce some concepts.

\subsection{$D=1$}
\label{sec_1D}

We can rightly ask if $D=1$ has any experimental relevance. The answer is positive
because a 2D vicinal surface may form an effective 1D pattern, 
when steps meander in-phase while preserving their distance: 
see Fig.~1(b1) for a clear experimental realization and Fig.~3b for its
theoretical modelization.
If this is the case, a rigorous analysis~\cite{Paulin2001,Misbah_review}  
has proved that the resulting step dynamics can be described
by an effective one-dimensional equation. Therefore, the 1D version of Eq.~(\ref{eq_2d}) describes the
dynamics of a vicinal surface if its steps syncronize along a high symmetry orientation.

According to a fairly general model~\cite{Paulin2001} which includes elastic
interactions among steps,\footnote{In this homoepitaxial context,
elastic interactions between steps or adatoms do not imply
long range interactions.}
the evolution of an
in-phase train of steps is ruled by the equation
\be
\p_t z = -\p_x \{ B(m) +G(m) \p_x [C(m)\p_x m ] \} .
\label{eq_1dgrowth}
\ee
If we change variable, defining
\be
u = \int_0^m ds C(s)
\ee
the evolution of $u$ satisfies
\be
\p_t u = - C(u)\p_{xx} [ B(u) + G(u)u_{xx} ],
\label{gCH}
\ee
which looks like a generalized Cahn-Hilliard (CH) equation.
In fact, the standard CH eq. is obtained if $C(u)$ and $G(u)$ are
positive constants and $B(u)=c_1 u - c_2 u^3$, with $c_{1,2}>0$. Its generalization, however,
should preserve the linear stability analysis of the trivial solution
$u\equiv 0$, which corresponds to say the functions $B,G,C$ have a fixed form for small $u$,
\be
B(u)\simeq c_1 u, \qquad G(u)\simeq c_3, \qquad C(u)\simeq c_4,
\ee
with all $c_i>0$. 
Given these relations, $u$ and $m$ are just proportional for small slope, $u\simeq c_4 m$.
Furthermore, if we assume $u(x,t)=u_0 e^{\sigma t + iqx}$ in the limit $u_0\to 0$ we get
\be
\sigma (q) = c_1 c_4 q^2 - c_3 c_4 q^4 ,
\label{eq_c1234}
\ee
which is the linear spectrum for the (generalized) CH equation.

The key point of our analysis is to study coarsening as a long-wave instability~\cite{Whitham} of
periodic steady states. This means that a periodic configuration may be unstable under large
scale fluctuations of the wavelength. In the case of Eq.~(\ref{gCH}), if we require that the constant 
average spatial value of the order parameter vanishes, $\langle u\rangle =0$ (which corresponds to
say the step has a high symmetry orientation) periodic steady states are given by the
vanishing of the quantity in square brackets, i.e. $u_{xx}= - B(u)/G(u)$.
Therefore, steady states correspond to the oscillations of a fictitious particle 
in the potential $V(u)=\int du B(u)/G(u)$,
which has the parabolic form $V(u)\simeq (c_1/2c_3)u^2$ for small $u$, i.e. for small $m$.
We leave out all technical details which can be found elsewhere~\cite{PRE_coarsening} and focus
on the spirit of the analytical treatment and, of course, on the main results.

\begin{figure}
\begin{center}
\includegraphics[width=0.9\textwidth]{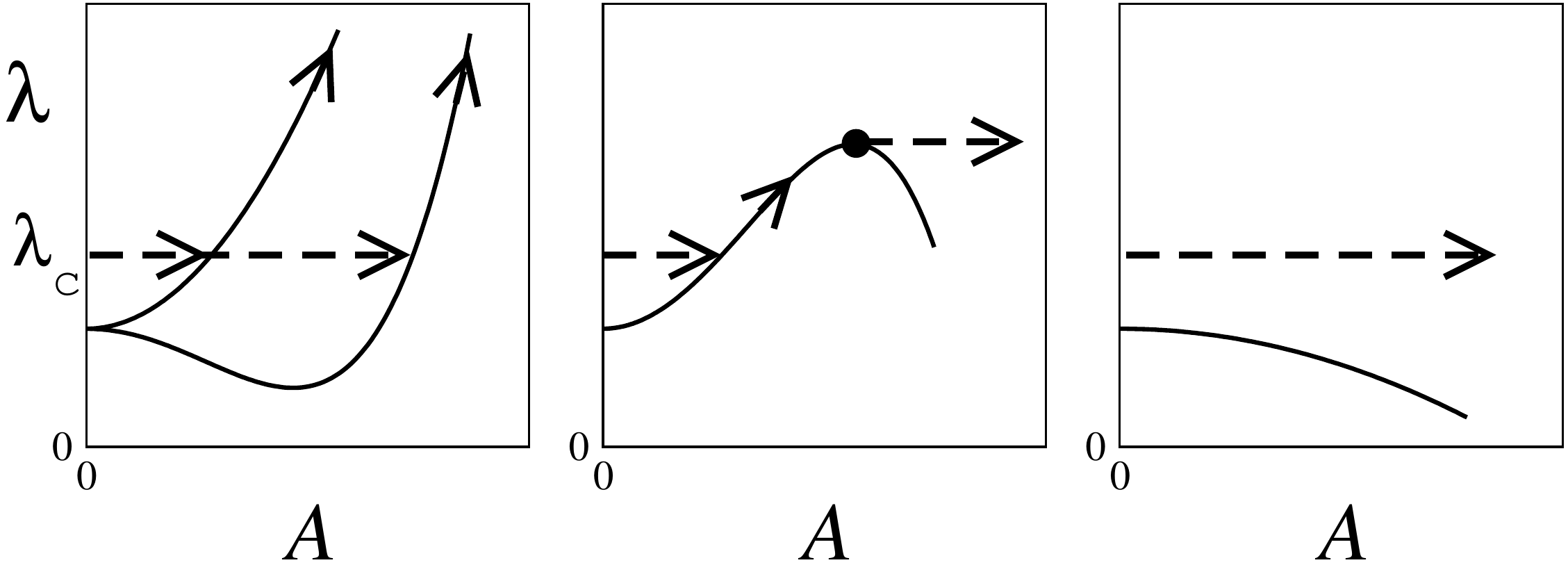}
\end{center}
\caption{Scenarios of perpetual (left), interrupted (centre), and no coarsening (right).
In all cases $\lambda(A\to 0)=\lambda_0$ is the minimal unstable wavelength when studying the
linear stability of the flat profile $u=0$. More precisely, if 
$\sigma(q)=c_1c_4 q^2 -c_3 c_4 q^4$ is the linear spectrum (see Eq.~(\ref{eq_c1234}))
then $\sigma(2\pi/\lambda_0)=0$. All wavelenghts larger than $\lambda_0$ are unstable and
$\lambda_c=\sqrt{2}\lambda_0$ is the most unstable one, since $\sigma'(2\pi/\lambda_c)=0$.
Arrows indicate dynamics in a qualitative way. Arrows ending dashed lines mean the pattern
has a constant wavelength and an increasing amplitude: this may occur either in the linear
regime (short times, small $A$) or at large times.
Arrows along a $\lambda(A)$ curve mean coarsening.
}
\label{fig_scenario}
\end{figure}

We need to study how a periodic steady state $u_\lambda(x)$ responds to a weak spatial perturbation.
Because of translational invariance, also $u_\lambda(x + \Phi)$ is a good steady state
and if the phase $\Phi$ acquires a slow spatial dependence, the temporal response is slow as well.
Formally, this amounts to assume that $\Phi=\Phi(X,T)$, where $X=\epsilon x$ and $T=\epsilon^2 t$
are slow variables and the different dependence on $\epsilon$ is attributable to the expected
diffusive character of the phase. The introduction of these new variables allows us to perform a
perturbative multiscale analysis, which finally provides an equation describing the
dynamics through the diffusion equation $\p_T\Phi = D\p_{XX}\Phi$, where 
the phase diffusion coefficient $D$ can be explicitly derived and be shown to have the form
\be
D = -D_0(A) \left(\frac{\p\lambda}{\p A}\right)^{-1}.
\ee
Here above $A$ and $\lambda$ are the amplitude and the wavelength of the steady states.
The quantity $\lambda$
(corresponding to the period of oscillation in the potential $V(u)$) is itself a function
of $A$. Since $D_0 > 0$, a phase instability ($D<0$) is present if and only if $\lambda$
is an increasing function of $A$. Therefore, we have a pretty simple criterion to 
discriminate among different scenarios, see Fig.~\ref{fig_scenario}: 
perpetual coarsening (increasing $\lambda(A)$ or $\lambda(A)$ has a minimum),
interrupted coarsening ($\lambda(A)$ has a maximum), and no coarsening (decreasing $\lambda(A)$).
In fact, we have even more than that, because the knowledge of $D(\lambda)$ allows to 
determine the coarsening law $\lambda(t)$ at large space and time scales.
In this regime the only relevant time scale is time $t$ and the only relevant
spatial scale is the wavelength $\lambda$ of the structure, so that
\be
|D(\lambda)| \approx  \frac{\lambda^2}{t} .
\ee

Using this relation and evaluating the asymptotic profile of steady states it is possible to
derive the coarsening exponent, $\lambda(t)\approx t^n$.
Using the explicit expressions of functions $B(u)$, $C(u)$, and $G(u)$ we can find 
when perpetual coarsening occurs and what is the coarsening exponent.

In practice, we have found that
coarsening is switched on by elastic interaction and $n=1/4$  or $n=1/6$, depending what type
of adatom diffusion is more efficient to relax the surface: $n=1/4$ if adatoms mainly diffuse along the
steps and $n=1/6$ if they diffuse mainly on terraces.
We should also add that the criterion of a maximum of 
$\lambda(A)$ signaling interrupted coarsening is able to explain the role of anisotropy~\cite{interrupted},
which has precisely this effect.
The theory has proved to be extremely powerful. Unfortunately,
from an experimental point of view we are not aware of meandering systems producing coarsening 
during all the experimental observation time: either coarsening is absent or it lasts very little.

\subsection{$D=2$}

Real two-dimensional, high-symmetry surfaces growing under slow deposition have been well studied
and coarsening is a common dynamical scenario, see Figs.~1a,c. 
However, we should start by saying that a quantitative comparison with experiments can be difficult
because attaining the asymptotic regime for enough time to allow the extraction of the
coarsening exponent $n$ from a
log-log plot of $L$ (the size of the emergent structure) vs $t$ is very hard.
In this respect we can observe that an exponent $n\le 1/3$ 
requires to increase time by at least a factor $10^3$ to increase $L$ by a factor $10$,
and this may not be feasible.

Performing a phase stability analysis in two spatial dimensions is by far more complicated
than in $D=1$, because a periodic structure is now defined by two wavevectors, so two phases
$\Phi_1,\Phi_2$ appear, each one depending on both spatial coordinates, $X_1,X_2$.
Therefore a diffusive behavior is now described by a system of equations,
\be
\p_T \Phi_i =\sum_{j,k,\ell} D^{ij}_{k\l} \p_{X_k X_\l}\Phi_j
\ee
which in principle means 12 different coefficients, even if the symmetries of the pattern
may strongly reduce such number. 

Our analysis has focused on the following class of equations,
\be
\frac{\p h(\mb{x},t)}{\p t} = -\nabla\cdot
[ \mb{j}(\nabla h) +\nabla(\nabla^2 h) ] .
\label{eq_growth}
\ee
If the current $\mb{j}$ is derivable from a potential, i.e. $\mb{j}= -\nabla_\mb{m} V(\mb{m})$, 
then taking the gradient of left and right hand sides of Eq.~(\ref{eq_growth}), the equation seems to reduce 
to a standard phase separation process,
\be 
\frac{\p \mb{m}}{\p t} = -\nabla \left\{ \nabla\cdot
[ \nabla^2 \mb{m} - \nabla_\mb{m} V(\mb{m}) ] \right\} .
\ee
However, this similarity is only apparent. Firstly, on the right hand side
we have $\nabla \nabla$ rather than $\nabla^2$: $\nabla \nabla$ means to take the gradient of the divergence,
while $\nabla^2$ means to take the Laplacian of each separate component. 
Secondly, $\mb{m}$ is a {\it constrained} order parameter, because
$\nabla\times\mb{m}= \nabla\times\nabla h\equiv 0$.
This constraint is related to the domain wall straightness:
if the slope is the order parameter two regions 
of constant, different slopes $\mb{m}_1,\mb{m}_2$, must forcely cross along a straigth line, so domain walls
are straight, which is not the case in standard phase separation processes. 
It is worth stressing that both remarks do {\it not} apply in $D=1$, where
crystal growth and phase separation do have strong similarities,
as evidenced by the generalized CH equation~(\ref{gCH}).

In $D=2$ instead, crystal growth dynamics as given by Eq.~(\ref{eq_growth})
are genuine new dynamics. We refer to papers~\cite{PRL_Sofia,PRE_Sofia} for
details and for a thorough discussion of the relevant bibliography. 
Here we report the main results focusing on the coarsening exponents
for models providing a perpetual coarsening scenario.\footnote{In $D=2$, the criterion
because coarsening occurs is a generalized version of the condition
$\p_A\lambda >0$, where the amplitude $A$ of the steady pattern is replaced by
a more complicated quantity. This quantity reduces to $A$ for small $A$.}
While the symmetry of the pattern does not seem to be a relevant feature,
the key point is the long time behavior of the slope in ``facets'',
which can either be constant (both in space and time) or can tend to increase
in time. This property can be established a priori via a simple inspection of
the current $\mb{j}(\mb{m})$, which can vanish for $\mb{m}=0$ only or also
for some finite values $\mb{m}_i^*$, called magic slopes and related by 
suitable rotational symmetry operations. If $\mb{j}= -\nabla_\mb{m} V(\mb{m})$,
these slopes are the minima of the potential.
According to our results, two main universality classes exist,
depending on the presence or the absence of magic slopes:
in the former case (faceting), $n=1/3$ while in the latter case
(increasing slope), $n=1/4$.

In the literature there have been several heuristic derivation of coarsening exponents
for crystal growth.
Here we report a simple explanation of the exponent $n=1/3$ in the case of faceting,
proposed by Golubovi\'c et al. in Ref.~\cite{Golubovic_review}.
Two hypotheses are used in such derivation, the first being the existence of one single length scale,
the wavelength of the periodic pattern. The second hypothesis is that
our model, Eq.~(\ref{eq_growth}), can be rewritten is the variational form
$\p_t h = -\delta {\cal F}/\delta h$, where ${\cal F}=\int d\mb{x} [
\frac{1}{2}\mb{m}^2 +  V(\mb{m}) ]$. Within these hypotheses, the excess free energy is concentrated
along domain walls, where the slope varies from a minimum of $V(\mb{m})$ to
another minimum, so ${\cal F} = \sigma\ss{DW}{\cal L}$, where 
$\sigma\ss{DW}$ is the domain wall (DW) energy cost per unit length and ${\cal L}$
is the total length of domain walls. If $\lambda$ is the average size of 
domains, there will be a DW length of order $\lambda$ in a region of order
$\lambda^2$, so ${\cal F}/{\cal A}\ss{s}\approx \sigma\ss{DW}/\lambda$, 
where ${\cal A}\ss{s}$ is just
the total surface area.
From the variational form of the evolution equation we can easily write that
\be
\frac{1}{{\cal A}\ss{s}} \frac{d{\cal F}}{dt} 
=\frac{1}{{\cal A}\ss{s}} \int d\mb{x} \frac{\delta {\cal F}}{\delta h}
\frac{\p h}{\p t}
=- \frac{1}{{\cal A}\ss{s}}\int d\mb{x} \left(\frac{\p h}{\p t}\right)^2 =-\langle (\p_t h)^2\rangle.
\ee
Because of faceting the height of the structure scales as $\lambda$, so that
$\langle (\p_t h)^2\rangle \approx (\p_t \lambda)^2$ and finally
\be
- (\p_t \lambda)^2 \approx \frac{1}{{\cal A}\ss{s}} \frac{d{\cal F}}{dt} =
\frac{d}{dt} \frac{\sigma\ss{DW}}{\lambda} ,
\ee
which gives $\lambda(t) \approx t^n$ with $n=1/3$.

In conclusion, when faceting occurs we expect $n=1/3$ if dynamics can be characterized 
by a single length scale, $\lambda$. This result is in agreement with our formal
derivation using the phase diffusion approach, but it is interesting to focus on the
hypothesis of a single length scale. According to our approach~\cite{PRE_Sofia},
the periodic pattern is unstable against long-wave fluctuations whatever is its symmetry.
However, when the current has a square symmetry~\cite{Levandovsky_Golubovic}
numerical simulations show a very irregular pattern which is characterized by two types
of domain walls, therefore by two length scales. In poor words, one length scale is the
length scale appearing in a perfectly periodic pattern, while the second length scale
measures the size of ordered regions, i.e. the distance between defects.
According to Refs.~\cite{Levandovsky_Golubovic,Siegert}
the structure would be metastable in the absence of such defects and the coarsening
may be slowed, giving $n=1/4$ instead that $n=1/3$.
While our results contradict the metastable character of the periodic array,
we cannot exclude that a {\it real} disordered pattern has a slower coarsening.
A similar question seems to arise for rectangular symmetry.
We refer the reader to~\cite{PRE_Sofia} for a more detailed comparison with
the literature.

\section{Elasticity}
\label{sec_el}

In the previous Section we have considered a growth phenomenon where
coarsening is a pure nonequilibrium effect: if we switch off the flux
coarsening stops and the surface relax towards the equilibrium state,
corresponding to a flat surface plus thermal roughening (even if this
relaxation may take a very long time).
In fact, the destabilizing term in the evolution equation, the current 
$\mb{j}$, is proportional to the flux~\cite{myCRP}. Furthermore, even if dynamics
is driven by the minimization of a suitable functional ${\cal F}$,
this pseudo free energy is a non equilibrium functional, whose form
is completely different from the true free energy driving the relaxation
towards equilibrium.

In this Section we are going to consider a problem which is apparently similar,
but whose phenomenology and theoretical description are fairly different.
The reason for that is {\it elasticity}, which comes into play when a solid
is grown epitaxially on top of a substrate of different material.
It is easy to understand what that means, because the ``epitaxial constraint",
forcing to growth an adsorbate with a lattice constant imposed by the substrate has an effect
similar to squeeze (or to stretch) a solid. Therefore, the system tries to reduce stress.
This may be done either {\it breaking} the epitaxial constraint or not.
The former case means that dislocations (i.e. defects breaking
the translational invariance) form. At the end dislocations always form, but we are rather
interested to the latter case, when epitaxy is maintained.
In this case, surface modulations form giving rise to the so-called 
Asaro-Tiller-Grinfeld (ATG) instability~\cite{Politi_review}.
This is the primary instability leading to quantum dot formation, see images (d1-d4)
in Fig.~1.

The typical phenomenology of a process of heteroepitaxial growth includes the
formation of a wetting layer, its destabilization just described, then
the formation of islands. As a matter of fact, most of the interest towards
these systems is related to their potential use as quantum
dots, whose distribution should be as uniform as possible. In practice
deposition is stopped after a while and the system is annealed waiting for
its stabilization. This after-growth dynamics is much longer than the growth
process itself and coarsening is expected to play a major role, with islands
exchanging atoms to relax their energy.

This process closely resembles Ostwald ripening, where the average size of an ensemble of islands/clusters
increases in time, driven by the Gibbs-Thomson effect~\cite{Saito_book}. 
Let's spend a few words on this 
phenomenon, assuming that $d-$dimensional clusters exchange matter in a $d-$dimensional space 
because atoms detach/attach from clusters and diffuse from a high density to a low density region.
The equilibrium density of atoms in proximity of a cluster of radius $R$ has a curvature dependent
contribution proportional to $1/R$ (Gibbs-Thomson effect), so there will be a net current $J$
from a smaller cluster to a larger cluster which reinforces their asymmetry in size and leads to
an instability. Since the current is proportional to the density difference ($\sim R^{-1}$)
and inversely proportional to the distance between clusters ($\sim R$), $J\sim R^{-2}$.
If a cluster of size $R$ is depleted by a current $J\sim R^{-2}$, it takes a time of order $R^3$
to disappear, so we expect the typical size of survived clusters at time $t$ is $R\sim t^{1/3}$.

This picture certainly cannot be applied to the heteroepitaxial case under study, and there
is no simple picture or simple analytical treatment we can propose. 
This is due to two reasons: firstly, the system is intrinsically complicated, with 
different energy terms which are all relevant: elastic stress, surface energy, and
wetting interaction with the substrate; secondly, there are several (control) parameters
which strongly affect the dynamics: the temperature, which controls surface diffusion, nucleation,
intermixing; the misfit between substrate and adsorbate, which controls the strength of the elastic
instability; the quantity of deposited material: the system first wets the substrate, then forms
ondulations and coherent dots, finally forms dislocations.

A good example of experimental overview is Ref.~\cite{Ge_Si1999},
where authors study the annealing of Ge islands on Si(100).
With increasing $T$, they pass from observing a final stationary state to a coarsening process
to an intermixing process between Ge and Si.
And limiting ourselves to the coarsening regime, observed at $T\approx 600^\circ$C,
island shape changes with size.
Changes of shape during coarsening have also been observed in precipitates in solids~\cite{Dahmen1997}
and leads to multimodal distributions of the size of aggregates. 
This fact clearly makes the theoretical analysis more difficult.

From the point of view of this paper the main question is whether the final state of the evolution is a periodic array
of islands/dots or if we should expect an ongoing coarsening.
The question has energetic and kinetic aspects and even from a purely energetic point of view
it is not trivial at all. In fact the minimization of the full energy is practically not feasible, because of
the many terms entering in the energy, the nonlocal character of elastic interactions, 
and the variety of periodic arrays that should be taken into account in the minimization.
Nonetheless, some attempts have been done, e.g., by V.A. Shchukin and collaborators~%
\cite{Shchukin1995,Shchukin_review,Shchukin2003},
finding that arrays are at least metastable.
Experimental data in agreement with this result and providing a stable or metastable bimodal
distribution (pyramids and domes) 
have been found by Medeiros-Ribeiro and collaborators~\cite{Medeiros-Ribeiro}.
However, this picture has been opposed in Refs.~\cite{Ge_SiTersoff1998,Rastelli},
where authors dispute the idea that different types of dots are in equilibrium with
each other. Rather, they suggest that a subtle interplay between thermodynamics and kinetics
is responsible of the dynamics of the system and the size distribution is
not steady but it continues to evolve through an anomalous coarsening involving a 
change of shape of dots.  

It follows that it is crucial to take into account kinetics 
through an evolution equation, but we should add that the resulting picture is strongly 
dependent on the basic ingredients used to build up the model. 
For example, linear elasticity and a non singular wetting potential have provided~\cite{Spencer2004}
a steady final state, but a different wetting interaction and its effect on stress make islands unstable and coarsening
is restored~\cite{Levine2007}.
A careful approach has been undertaken by J.-N. Aqua and collaborators, who have included nonlinear elastic
effects~\cite{Aqua2007,Aqua_review} and anisotropy effects~\cite{Aqua2013}.
According to their results the isotropic system displays interrupted coarsening, while anisotropy
changes islands' shape (as expected) and arrest coarsening. This picture of interrupted coarsening
is also found experimentally~\cite{Aqua2013} on SiGe/Si, see Figs.~1d.

\section{Erosion}
\label{sec_er}

So far we have discussed coarsening phenomena having two common features:
they are driven by thermally activated elementary processes and they occur during
a growth process (or during the annealing following a growth process).
This Section differs in both aspects, since the morphology is driven by bombarding
and eroding the surface with an energetic flux of ions.
Each ion penetrates the surface and releases its energy in the surrounding environment,
which allows to kick atoms out.\footnote{Sputtered atoms can then deposit onto another surface. 
We are not interested here in this amorphous growth process, but in the erosion process.}

In Fig.~1(e1-e4) we reproduce some STM topographs by Martin Engler et al.~\cite{Engler} showing
a Si(001) surface after Kr$^+$ bombardment: the initially flat surface is destabilized
and an array of ripples appears with an increasing wavelength.
Such a regular morphology (attested by height profiles, see~\cite{Engler})  
is not obvious and we might expect that eroded surfaces undergo a kinetic roughening process, 
as it actually happens in other cases~\cite{Eklund1991}.
It is hard to evaluate a priori the relevance of noise and in the literature there are 
analyses of sputter erosion based on stochastic equations like the anisotropic 
Kardar-Parisi-Zhang equation~%
\cite{Cuerno_Barabasi}. However, most studies are focused on deterministic equations where 
roughening is the result of an instability. We will limit here to deterministic processes.

The primary question when facing an instability concerns its origin: why does it rise?
The first answer was proposed by Bradley and Harper~\cite{Bradley_Harper} who suggested that erosion 
velocity might depend on the curvature, being larger for a positive curvature. This fact alone leads to an
evolution equation like $\p_t h = -\nu\nabla^2 h$, which displays an instability.
In subsequent years other mechanisms and linear theories have been proposed, evoking 
mass redistribution~\cite{mass_redistribution}, ion induced flows~\cite{ion_induced_flow}, and so on.
These few words just to make it clear that physical processes at the origin of the instability are
not fully understood, also because they may depend on the material and on the erosion conditions.

A linear theory is also able to address the size and the orientation of the emergent structure
and different answers can be obtained to questions about ripple rotation and the effective unstable
character of the flat surface with varying the incidence angle of the beam.
For example, recent experimental answers to these issues have appeared for
erosion patterns on Ge(100)~\cite{Teichmann} and Si(001)~\cite{Engler}.
In these papers authors do not observe ripple rotation and they don't even observe
ripple formation for normal or quasi normal incidence. Instead, for oblique incidence 
they record coarsening processes, see Figs.~1e for Si. 
Coarsening is described as a nonlocal process,
due to the reflection of ions from the downwind face towards the upwind face~\cite{Hauffe}.

A different coarsening picture emerges from other erosion experiments on 
Si~\cite{Garcia2010,Garcia2012,review_Cuerno}, where
authors report to observe interrupted coarsening. 
It is possible, as the authors say, that metal impurities and silicide formation could influence the pattern morphology
and its dynamical evolution. 
Their picture of interrupted coarsening is based on an analytical model which is worth discussing,
\be
\frac{\p h}{\p t} = -\nu\nabla^2 h - {\cal K}\nabla^4 h + \lambda_1 (\nabla h)^2 
-\lambda_2 \nabla^2 (\nabla h)^2 .
\label{eq_Cuerno}
\ee

The linear part is standard: the $\nu-$term is responsible for the instability and the ${\cal K}-$term
heals the instability at short lenght scales. Their balance produces an initial pattern
of wavelength $\lambda\ss{c}=2\pi(2{\cal K}/\nu)^{1/2}$.
The nonlinear part is composed of a conserved term ($\lambda_2$) and a nonconserved one ($\lambda_1$).
Without the latter term, the above equation becomes the conserved Kuramoto-Sivashinsky equation,
$\p_t h = -\nabla^2 (\nu h + {\cal K}\nabla^2 h + \lambda_2 (\nabla h)^2 )$,
which is known to display perpetual coarsening~\cite{Frisch_Verga,Matteo}.
In the opposite limit, $\lambda_2=0$, we get the standard Kuramoto-Sivashinsky equation,
which is known to display spatio-temporal chaos with a cellular structure of order $\lambda\ss{c}$.
It is not trivial that their combination may produce interrupted coarsening
and it would be interesting (but fairly complicated in 2D) to study Eq.~(\ref{eq_Cuerno}) 
with the phase diffusion approach discussed in Sec.~\ref{sec_hs}.

\section{Final considerations}

The first general remark is that dynamics may be driven by some free energy. 
It may be a real free energy, see Sec.~\ref{sec_el}, or a pseudo free energy, see Sec.~\ref{sec_hs}.
It is not granted that dynamics allows to attain the ground state and the interruption of
coarsening may be determined by the trapping in some metastable state. The presence of elasticity
makes the problem so complicated it is not even possible to know what pattern minimizes the energy,
especially in two spatial dimensions, which is the only physically relevant case.
Furthermore, simple models of the ATG instability produce finite time singularities
which need to be regularized~\cite{regularization_ATG}.

In the kinetic-driven case, Sec.~\ref{sec_hs}, the existence of a pseudo free energy is not really relevant.
For example, the one-dimensional problem describing the in-phase meandering of a vicinal surface
is studied by Eq.~(\ref{eq_1dgrowth}) which cannot always be put into a variational form. In spite of this,
its analysis by the phase diffusion equation can be performed giving a very general criterion for the existence 
of coarsening: the wavelength of periodic steady states is an increasing function of their amplitude.
The same approach can be employed in 2D, even if such simple criterion seems to be applicable only for
small amplitude structures, in the general case the coarsening criterion being the negativity
of a suitable diffusion coefficient.

Apart from the exact coarsening criterion, an important feature of the kinetic-driven case is that
interrupted coarsening is accompanied by a diverging amplitude, the final state as described by
Eqs.~(\ref{eq_1dgrowth},\ref{gCH}) not being stationary. This scenario corresponds to what is really 
observed, e.g., in the growth of Pt/Pt(111), where wedding cakes of increasing height and constant 
lateral size are finally obtained, see Figs.~1c. This system seems to be the experimental realization of the simple
Zeno model introduced long time ago~\cite{Elkinani,Politi_Villain}.

In the other cases studied here, interrupted coarsening means attaining a final structure which is finite
in wavelength {\it and} amplitude, even if it may not correspond to the minimal energy state.
In general terms, it seems that at least three mechanisms of interrupted coarsening exist:
the simplest one occurs when dynamics is driven by a free energy which is minimized by a structure
of finite size; the second one corresponds to the kinetic-driven case of our Sec.~\ref{sec_hs},
where the final state has a diverging amplitude; the third one appears in the erosion problem, 
Sec.~\ref{sec_er}, where the interruption seems to be the result of a competition between a perpetual
coarsening dynamics and a constant wavelength scenario.
This classification does not pretend to go beyond the problems studied here. For example, it is known that
quenched disorder~\cite{Federico} or the existence of stirring in a binary fluid may also lead to coarsening interruption
or to its sharp slowdown. 

When a perpetual coarsening process occurs, it is important to study its growth law, usually a power law one.
The phase diffusion approach allows to determine the coarsening exponent $n$, but our results for the 2D growth
law on a high symmetry surface do not agree with other findings, especially numerics, concerning square symmetry.
A possible explanation is that the emerging square pattern seems to be fairly disordered, so that the
time scale derived from a periodic array might not be the appropriate one.

The topics discussed in this paper have the nice feature to be characterized by a strong
link among theory, simulations and experiments. On the one hand, the more a phenomenon is widespread the
more it is worth studying. On the other hand, it would be important to have an agreed ``model system"
so as to allow a quantitative comparison between theory/simulations and experiments.
Coarsening dynamics are widespread in each of the three subfields discussed here,
but it seems to us that some efforts are needed for the energetic and the athermal cases
in order to identify a {\it simple} theoretical model,
or a {\it simple} experimental system, or both. Nobody can assure this is possible, of course.

\section*{Acknowledgements}
I acknowledge Isabelle Berbezier for the experimental images (d1-d4) in Fig.~1.
I profited a lot from discussions and from the critical reading of the manuscript of several persons:
Jean-No\"el Aqua, Claudia Innocenti, Thomas Michely, Chaouqi Misbah, Matteo Nicoli, and Jacques Villain.

\bibliographystyle{phaip}


\end{document}